# Knowledge-Reuse Transfer Learning Methods in Molecular and Material Science

An Chen[1,2,†], Zhilong Wang[1,2,†], Karl Luigi Loza Vidaurre[1,2], Yanqiang Han[1,2], Simin Ye[1,2], Kehao Tao[1,2], Shiwei Wang[1,2], Jing Gao[1,2], and Jinjin Li[1,2*]

[1]National Key Laboratory of Advanced Micro and Nano Manufacture Technology, Shanghai Jiao Tong University, Shanghai, 200240, China

[2]Department of Micro/Nano Electronics, School of Electronic Information and Electrical Engineering, Shanghai Jiao Tong University, Shanghai, 200240, China

†These two authors contribute equally to this study.

*Correspondence to: Jinjin Li (lijinjin@sjtu.edu.cn)

# Abstract

Molecules and materials are the foundation for the development of modern advanced industries such as energy storage systems and semiconductor devices. However, traditional trial-and-error methods or theoretical calculations are highly resource-intensive, and extremely long R&D (Research and Development) periods cannot meet the urgent need for molecules/materials in industrial development. Machine learning (ML) methods based on big data are expected to break this dilemma. However, the difficulty in constructing large-scale datasets of new molecules/materials due to the high cost of data acquisition and annotation limits the development of machine learning. The application of transfer learning lowers the data requirements for model training, which makes transfer learning stand out in researches addressing data quality issues. In this review, we summarize recent advances in transfer learning related to molecular and materials science. We focus on the application of transfer learning methods for the discovery of advanced molecules/materials, particularly, the construction of transfer learning frameworks for different systems, and how transfer learning can enhance the performance of models. In addition, the challenges of transfer learning are also discussed.

# 1. Introduction

Growing demand for technology and industry has led to a diverse suite of advances in molecular and material science. Current computational techniques, such as quantum mechanical (QM) methods, have become increasingly essential for interpreting experimental phenomena and providing the mechanistic understanding necessary to design more effective molecules and materials[1-8]. However, high-precision computational techniques incur high expenses; As the number of atoms in the system increases, the cost of computing increases exponentially. It is urgent to develop more efficient methods under the condition that the predictive accuracy is equal to the traditional computational methods. Machine learning (ML) is the branch of artificial intelligence (AI) that deals with exploratory tasks through feature engineering and optimization methods. In general, we categorize ML into supervised learning (SL) and unsupervised learning (UL) based on the scale of labeled data available. With the craze of "AI for Science", both SL and UL methods have begun to break new frontiers in many fields, including biology[9-11], physics[12, 13], and chemistry[14-19]. This is exemplified by the recent triumph of AlphaGo[20], AlphaFold[21] and AlphaMat[22].

Machine learning (ML) techniques form an important hub that drives databases, attributes, and applications of molecules and materials, bringing molecule and material design to the next stage of rapid development. Nevertheless, the application and implementation of traditional SL methods in molecule and material design are challenging, and physical insights from large databases with material attributes remain limited owing to the substantial engineering requirements of data resources. Training methods based on small sample data have become the direction of current research. UL methods, which do not require data annotation, have attracted attention in the field of molecule and material discovery in recent years[23-25]. However, it is precisely due to the lack of data annotations that UL cannot construct accurate structure-activity relationships, making it difficult to understand the mechanism of molecules and materials from a physical and chemical perspective.

Transfer learning (TL) is the revolutionary method of SL, which possesses the advantages of SL and UL. Namely, TL can greatly reduce the cost of data annotations, and construct accurate structure-activity relationships, which can achieve great prospects in molecular and materials science[26-30]. TL models adopt a strategy that can recognize and apply knowledge learned from source domain/tasks to target domain/tasks. An important factor that promotes the universality of TL is its reuse of the data of the existing source domain/tasks; therefore, it does not need to incur the high costs of compiling massive new databases. In our previous work, we are the pioneers of proposing the material modeling concepts of "horizontal transfer"[31, 32] and "vertical transfer"[8, 33], providing effective guidelines for material modeling requirements. Horizontal transfer is to reuse chemical knowledge in different material systems, while vertical transfer is to reuse chemical knowledge in different data fidelity of the same material system. For instance, Wang et al.[31] proposed a method that can efficiently predict the adsorption capacity of adsorbents at arbitrary sites based on a horizontal transfer strategy, which compensates for the lack of active site and adsorption data, and allows the model to be transferred to an arbitrary material using small amount of data (~10% of the data required to train the high accuracy model in general), and the final adsorption energy prediction model has a root-mean-square error (RMSE) of 0.1 eV. The construction of a high-precision, low-cost database of macromolecule (such as biomacromolecule) is extremely expensive. Han *et al.* proposed a vertical transfer strategy that can be used to take the low-precision data of the system as the source field, and optimize the target field through transfer learning, i.e., to obtain high-precision force field data, which reduces the amount of high-quality data required to about 5% of that of the general method[33]. Therefore, TL offers a solution to the molecular and material exploration problem, making predictions of new molecules and materials from small datasets, which in turn can drive the generation of more data that can be used to further refine the ML models.

Here, we will provide an in-depth review of knowledge-reused TL methods in molecular and material science, a field where new molecules and materials with high performance (e.g., suitable electronic properties, strong adsorption energy, etc.) can have a transformative impact on the urgent problems of physics and chemistry. The next four sections will provide a concise overview of TL concepts, including the definitions of TL, categorization of TL and the key problems that TL can address, with the goal of giving the reader an appreciation of state-of-the-art techniques as well as resources for building TL models for molecules and materials. Then, section VI reviews the practical application of TL techniques to the models in molecular and material science (e.g., experimental characterization analysis, macroscopic physical properties, reaction thermochemistry, isomerization, drug-like molecular torsions, formation energy, electronic conductivity, etc.). The final section outlines our perspective on the challenges and opportunities in TL for molecule and material design.

## 2. Nuts and bolts of transfer learning

### 2.1 What is transfer learning

As early as the end of the 20$^{th}$ century, TL attracted more and more attention under different terms, such as lifelong learning, knowledge transfer[29, 34], inductive transfer[35], multi-task learning[36, 37], meta-learning and incremental learning[38]. All of these emphasize that TL is a framework that can support multi-task learning. A typical approach to multi-tasking learning is to reveal common (underlying) features that each task benefits from. As shown in Fig. 1, traditional ML techniques focus on learning knowledge from each task independently, while TL methods focus on transferring knowledge from previous tasks to target tasks when the target tasks lack high-quality training data.

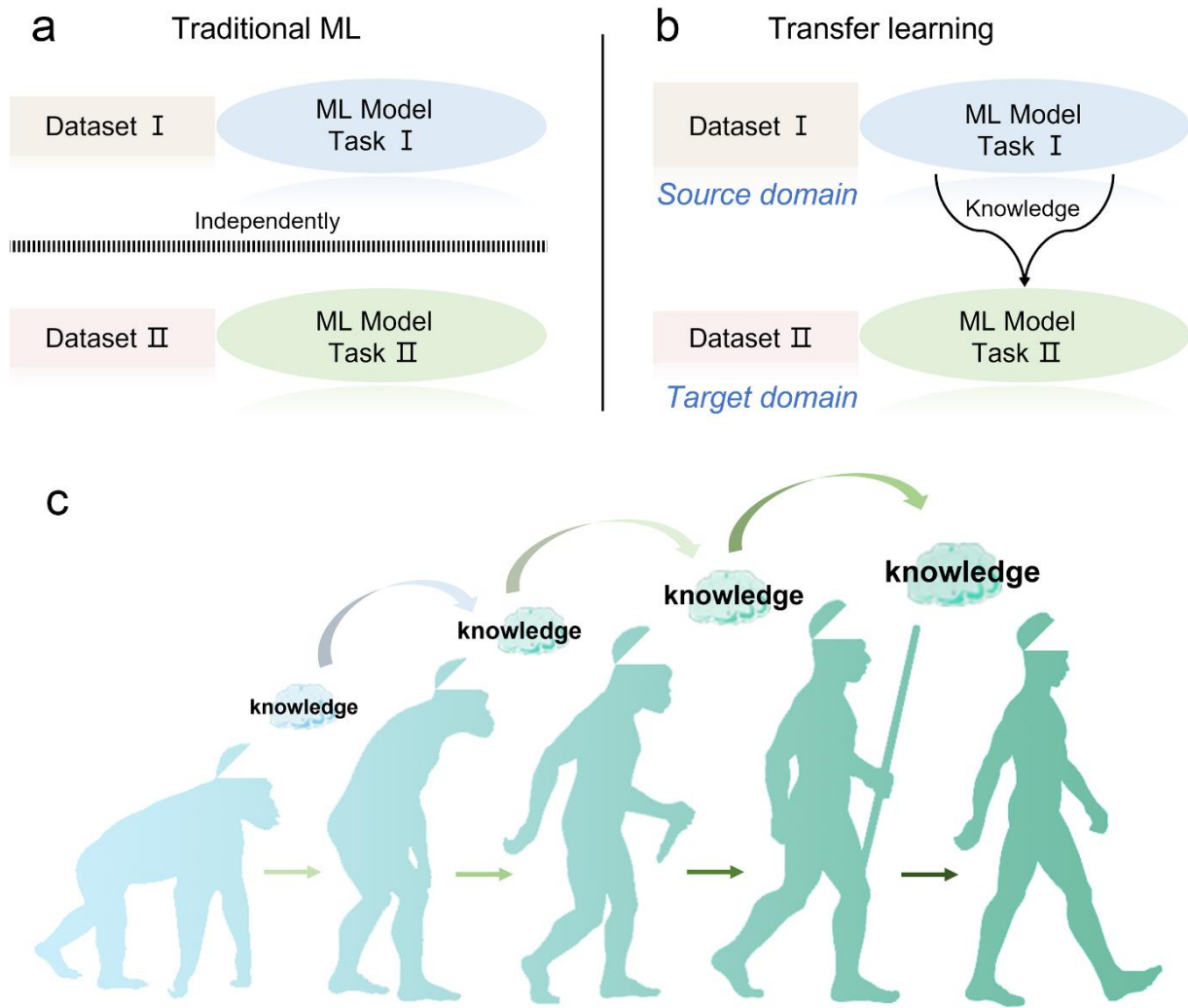

**Figure 1** Differences between traditional machine learning and transfer learning: (a) Traditional machine learning is usually a single learning task with no exchange of information between tasks. (b) Transfer learning focuses on the exchange of knowledge between different tasks and thus solves problems such as data scarcity. (c) Knowledge-reuse of transfer learning. TL is the cross-domain transfer of knowledge, which stems from a concept related to the generalization and transfer of human experience. That is, after generalization of experience, it becomes possible to transfer across multiple contexts to solve more problems, and the model (e.g., humans at various stages of life as plotted in the figure) will become more and more powerful.

In this section, we introduce the basic definitions and notations of ML and TL. In ML models, there are two basic definitions: domain ($D$) and task ($T$). The $D$ consists of two parts: a feature space $F$ and a marginal probability distribution $P(X)$, where $X = \{x_1, x_2, …, x_n\} \in F$. In the case of the goal of predicting whether the material is a metal or a semiconductor, $x_i$ is the $i$th feature vector related to the material's properties on electronic conductivity. Generally speaking, if two domains are different, they have different $F$ or obey different $P$. Given a $D = \{F, P(X)\}$, a $T$ is made by a label space $Y$ and a target prediction function $f(\cdot)$ ($T = \{Y, f(\cdot)\}$), where $f$ can be used for predicting the label for a new $x$, and can be thought of as $P(y|x)$ in probabilistic terms (conditional probability distribution). $T$ cannot generally be predicted intuitively, but can be learned from training data $\{x_i, y_i\}$, where $x_i \in X$, $y_i \in Y$. In this sense, if two learning tasks are different, it may be a different $D$ or a different $T$, and a different $D$ may be due to a different $F$ or $P(X)$, a different $T$ may be due to a different $Y$ or $f(\cdot)$[39].

When modeling using the TL methods, there is a source domain ($D_s$) and a source task ($T_s$), a target domain ($D_t$) and a target task ($T_t$), where $D_s \neq D_t$ or $T_s \neq T_t$. Then, TL is committed to using knowledge learned from the $D_s$ and the $T_s$ to help learn the $f(\cdot)$ in the $T_t$. In the case of the goal of predicting whether the material is a metal or a semiconductor, if the $D_s$ and $D_t$ are different, it may be because the predicted objects are small molecule organic materials and inorganic crystal materials, resulting in different feature spaces, or because different data sets of inorganic crystal materials are used, resulting in different marginal probability distributions; If the $T_s$ and $T_t$ are different, the label space may be different due to the different categories (binary: metal/semiconductor, ternary: metal/semiconductor/insulator) of conductive properties, or the $f(\cdot)$ may be different due to the different precision of conductive properties (experimental value and computational value) in the label space. As a result, many different modeling purposes or goals emerge as domains and tasks differ, and domains or tasks are considered related when there is a relationship between them, either explicitly or implicitly.

Instead of learning from scratch, utilizing transfer learning is based on models learned during previous problems encountered in a variety of ways, enabling accurate models to be built in a short period of time. Previous learning can be utilized, similar to standing on the shoulders of giants. From a practical point of view, transfer learning roughly consists of two processes: (1) pre-training; and (2) fine-tuning the model. It is worth noting that the difference between transfer learning and fine-tuning needs to be clarified. Fine-tuning is a specific method in transfer learning for adapting a model to the target task based on the source task. During the fine-tuning process, it is common to unfreeze some or all of the parameters of a pre-trained model and further train the model using the dataset of the target task. The steps of fine-tuning include: (1) freezing: locking the parameters of the pre-trained model to prevent them from being updated. (2) unfreezing: unlocking some or all of the parameters of the pre-trained model so that it can be trained for fine-tuning on data from the target task. The goal of fine-tuning is to adapt a pre-trained model to the features and data distribution of the target task through limited training on the target task. With fine-tuning, the model can be quickly adapted to achieve better performance with a small amount of target task data.

**2.2 Types of transfer learning**

According to the different situations of $D_s$ and $D_t$, $T_s$ and $T_t$, TL can be divided into three categories (as shown in Figure 2): (1) Inductive TL, where the $D_s$ and $D_t$ are the same, while the $T_s$ and $T_t$ are different but related. If the $D_s$ contains adequate annotated data, inductive TL is similar to multi-tasking learning. To be clear, inductive TL only focuses on obtaining higher model performance in $T_t$ by transferring knowledge from $T_s$, while multi-tasking learning tries to learn both $T_s$ and $T_t$. If there is no large amount of annotated data available in the $D_s$, inductive TL is similar to self-learning, that is, the label space of the $D_s$ and the $D_t$ may be different, which means that the marginal information in the $D_s$ cannot be directly used. (2) Transductive TL, as opposed to inductive TL, where the $T_s$ and $T_t$ are the same, while the $D_s$ and $D_t$ are different but related. In this case, the annotated data is scarce in $D_t$ and adequate in $D_s$. The difference in the $D_s$ may be due to the difference in the feature space or the difference in the marginal probability distribution of the input data. (3) Unsupervised TL, similar to inductive TL, $T_s$ and $T_t$ are different but related. However, unsupervised TL focuses on solving unsupervised learning problems in the $D_t$, such as clustering, dimension reduction, density estimation, etc. That is, no annotated data is available for the training set of both the $D_s$ and the $D_t$.

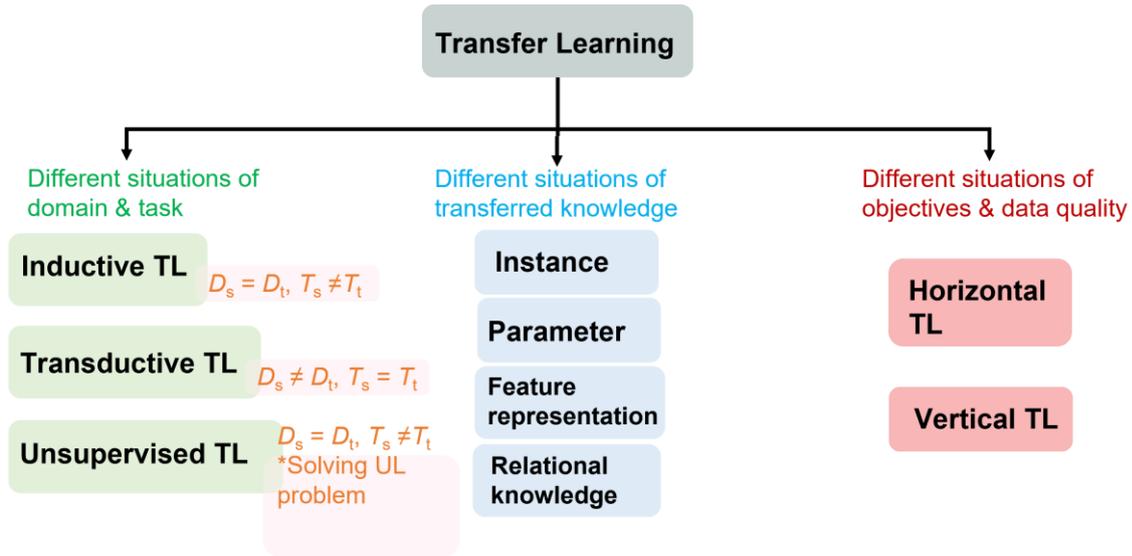

**Figure 2** Classification of transfer learning. Left: Based on the relationship between domain and task, it can be categorized as: Inductive TL, Transductive TL and Unsupervised TL; Center: Based on the transferred knowledge, it can be categorized as: Instance TL, Parameter TL, Feature Representation TL and Relational Knowledge TL. Right: Based on the research objectives and data quality, it can be categorized as Horizontal TL and Vertical TL.

According to the different situations of transferred knowledge, TL can be divided into four categories (as shown in Figure 2): (1) Transfer based on instance, which means that some annotated data in the $D_s$ can be reused by adjusting the weight, and then used for learning in the $D_t$ The adjustment of instance weight and importance sampling are the main techniques of this method. (2) Transfer based on feature representation, is to train a good representation for the $D_t$. In this case, the knowledge used for cross-domain transfer is encoded as the feature representation of learning to reduce the difference between the feature representation of the $D_s$ and the $D_t$, thus improve the performance of the $T_t$. (3) Transfer based on parameter, which assumes that Ts and Tt share some parameters or previous hyperparameter distributions of the model's hyperparameters. The transferred knowledge is encoded as shared parameters or priors. Thus, knowledge can be transferred across tasks by discovering shared parameters or priors. (4) Transfer based on relational knowledge, which is used to deal with related domains. It assumes that the relationship between some data in the $D_s$ and the $D_t$ is similar, so the transferred knowledge is the relationship between the data.

In addition, based on the research objectives and data quality, we propose a more intuitive

TL classification method that is more closely related to practical applications: (1) Horizontal TL, with different research objectives but the same data accuracy: use the data of one material/molecule and some existing materials/molecules as the based data of source domain model, and use other materials/molecules as the target domain. (2) Vertical TL, with the same research objective but different data accuracy: use the low accuracy property data of materials/molecules as the source domain, and optimize the target domain through transfer learning, i.e., obtain the high accuracy property data. This categorization helps researchers to quickly select a transfer learning strategy based on the research objectives and the key problems encountered, e.g., horizontal TL can be selected when the data on the target properties of the material/molecule are extremely scarce, and vertical TL can be selected when the data on the target properties of the material/molecule are lacking only in terms of accuracy.

**Table 1 Difference between transfer learning and other related learning algorithms.**

| Learning Algorithm | Training | Testing | Condition |
|---|---|---|---|
| Transfer learning | $D_s, D_t$ | $D_t$ | $T_s \neq T_t$ |
| Multi-task learning | $D^1, \ldots, D^n$ | $D^1, \ldots, D^n$ | $T_i \neq T_j, 1 \leq i \neq j \leq n$ |
| Domain adaptation | $D_s, D_t$ | $D_t$ | $P(X_s) \neq P(X_t)$ |
| Domain generalization | $D^1, \ldots, D^n$ | $D^{n+1}$ | $P_i \neq P_j, 1 \leq i \neq j \leq n$ |
| Meta-learning | $D^1, \ldots, D^n$ | $D^{n+1}$ | $T_i \neq T_j, 1 \leq i \neq j \leq n+1$ |
| Lifelong learning | $D^1, \ldots, D^n$ | $D^1, \ldots, D^n$ | *Sequential $D^i$* |

The boundaries between the three concepts of transfer learning and domain adaptation (DA) and domain generalization (DG) are often blurred. In general, transfer learning centers on accomplishing predictions in the target domain by reducing the distributional differences between the source and target domains, thereby utilizing the information learned in the source domain. Transfer learning trains a model on a source task with the goal of improving the model's performance on a different but related target domain/task. Pre-training-fine-tuning is a common strategy for transfer learning, where the source and target domains have different tasks and the target domain is visited during training. The aim of domain adaptation is to take data from source and target domains with different distributions and map them into a feature space where the data from the source and target domains are as similar as possible in the feature space. DA solves the problem of the difference in distributions that exists between source and target domains and maximizes the predictive performance of the model for a given target domain. The problem faced by DG is also the problem of the difference in distributions between the source and target domains, but unlike TL and DA, the target domain in scenarios using DG does not have any a priori information, i.e., DG does not see the target domain data during training, the target domain is inaccessible, and the training and testing tasks tend to be the same, but they have different distributions. However, DA has a small amount of target domain data at training time, which can be labeled or unlabeled. Therefore, DG is more challenging than DA, but more realistic and favorable in practical applications. Based on the similarities and differences of feature space and label space, DA can be categorized into homogeneous DA and heterogeneous DA, while DG can be categorized into single-source domain DG and multi-source domain DG by the type of source domains. The differences between DG and DA, and between them and multi-task learning, transfer learning, meta-learning, and lifelong learning are shown in Table 1. As can be seen from Table 1, meta-learning is a generalized learning strategy where the inputs are a large number of training tasks and corresponding training data, and the purpose of this strategy is Learning-to-Learn. Lifelong learning is concerned with the ability to learn between multiple consecutive domains/tasks, which requires that the model learns over time while retaining the information learned in the previous stage. All algorithms presented above are strategies or evolutions of transfer learning.

Distinguishing between various learning algorithms will help us to choose the right method in solving practical problems. Another concept that is often confused with TL is "pre-training". In fact, pre-training is the "carrier" of TL strategy, TL is usually manifested through the use of pre-trained models. A typical example is neural network algorithms based on TL. The use of transfer learning has changed the framework of traditional neural network model training, and the introduction of pre-training steps based on different domain datasets has led to a significant increase in the generalization of neural network models, as well as a significant reduction in the consumption of computational resources and an increase in efficiency. Deep Convolutional Neural Networks (DCNN) have excellent feature extraction capabilities and are therefore widely used as pre-training models. Table 2 shows the comparison between some widely used pre-training models.

**Table 2 Comparison between several pre-trained models.**

| Learning Model | Advantages | Weaknesses |
|---|---|---|
| GoogleNet[40] | *Implements a more efficient handling of non-linear activations.* | *As the number of layers increases, training becomes more difficult and might fail to converge.* |
| AlexNet[41] | *Fast training, low computational complexity, use of dropout, prevents overfitting.* | *The number of epochs required for network convergence is usually high.* |
| VGGNet[42] | *Simple but powerful structure with a high accuracy.* | *Increased number of parameters, larger memory usage, more computationally expensive.* |
| ResNet[43] | *The gradient vanishing problem is solved by using interlayer residual jump connections.* | *Prolonged training time and might suffer of redundancy.* |
| DenseNet[44] | *Robust against gradient vanishing, strong feature transfer, and reduced number of parameters.* | *Large amount of memory used during training.* |
| MobileNet[45] | *Reduced number of parameters and faster calculations.* | *Lower overall accuracy.* |

## 2.3 Why use transfer learning

Computational methods based on QM (*e.g.*, density functional theory (DFT)) are the most widely used in computational science of chemistry and materials[46-50]. In the process of computational simulation, massive and high dimensional data sets are usually generated[5, 51-53]. ML methods, such as decision trees (DT)[54], support vector machines (SVM)[55], random forests (RF)[56], neural networks (NN)[57], etc., can carry out in-depth mining of these data sets, identify the linear and nonlinear patterns, discover potential laws and trends in data that are difficult to be found by traditional statistical methods, and provide possibilities for property prediction, efficient screening and reverse design of molecules and materials. At present, ML methods have been widely used in drug molecule design[58, 59], biomacromolecule transformation[60, 61], and new material research[62-64]. For example, Russ *et al.* used an evolution-based algorithm for protein field construction, they reported a data-driven framework for learning the constraints of specifying proteins from evolutionary sequence data, and built a database for synthetic genes, testing their activity in vivo[65]. Wang et al. used an unsupervised learning algorithm for photovoltaic semiconductor screening, and described an ensemble learning algorithm for battery material discovery, catalytic activity prediction[15, 66].

In these machine learning and data mining algorithms, an important assumption is that the training data in those works and the data collected for other tasks must have the same feature space and distribution; however, this assumption does not hold true in many practical applications. For example, researchers have enough data on the catalytic activity of different materials for more reactions, such as oxygen reduction reaction (ORR), oxygen evolution reaction (OER), $CO_2$ reduction reaction (CO2RR) and nitrogen reduction reaction (NRR). It is clear that the data generated by these catalytic reactions have a large feature space and the probability follows different data distributions. When such data distribution changes, most statistical models and traditional ML methods must collect new data samples for model reconstruction, which requires a large data compilation cost and cannot be realized in a short time. Then, if the need and cost of recollecting training data is reduced, the cost of spending a large number of labeled sample data can be avoided. In this case, the transfer of knowledge or TL between different molecule and material domains will become necessary.

## 3. Advances of TL methods in molecular and material science

ML approaches have achieved great success over the past decade due to increased data availability and improved algorithms. ML is increasingly used in experimental and computational chemistry. Many advances aim to revolutionize chemistry by applying computational science to chemical and biological systems. However, the current lack of data on molecules/materials in terms of quantity and breadth limits the usefulness of machine learning, and this situation is not likely to change in the near future. Therefore, the introduction of transfer learning can overcome this difficulty, using a small amount of data to predict the properties of molecules/materials, and improve the accuracy of the model. Practical applications of TL in molecular and material science are discussed in the following section, the importance and advantage of TL in these fields are also emphasized.

*Small molecules*

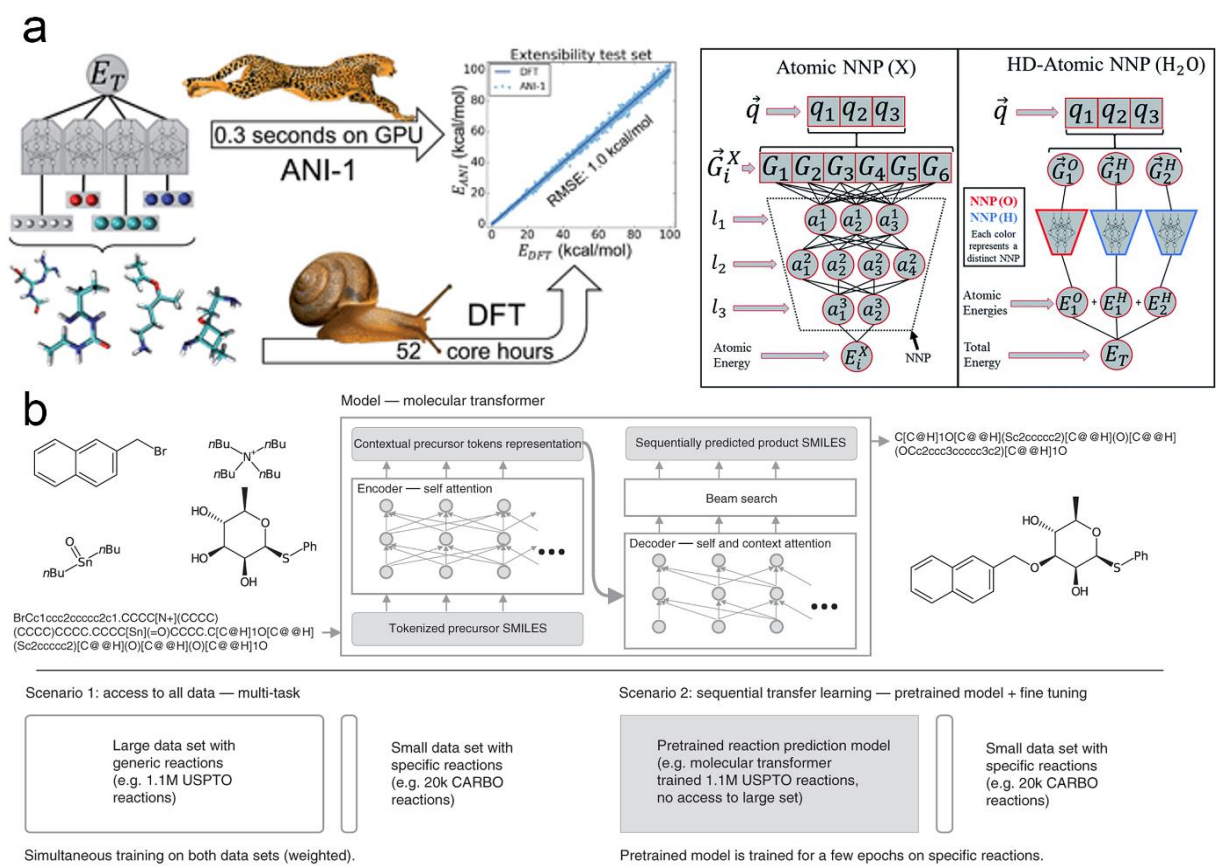

**Figure 3** (a) Work framework according to Behler and Parrinello's HDNN or HD-atomic NNP model. Reprint from Ref [67] (b) Molecular Transformer model and data scenarios. Reprint from Ref [73].

Since molecular dynamic simulation is an important foundation and key means to understand chemical and biological structures, the transferability, accuracy, and fast prediction of the molecular energy and atomic forces are particularly important for the next generation of linear-scaling model potential energy surface.

As one of the most widely used approaches to ML today, deep learning has been applied to many areas of science and technology, especially image, text and speech recognition, and recent fruits in areas such as chemistry and materials are noteworthy. However, often these deep learning models are non-transferable and cannot be applied, for example, to bulk materials and water cases. Moreover, when using deep learning techniques, the number of parameters in the models increases significantly, requiring large amounts of training data. Among the researches of small molecule, constructing deep learning models is often impeded by insufficient data size and quality. Smith *et al.*[67] demonstrated how to effectively implement TL in molecular property prediction to develop high-dimensional neural network potentials (HDNNPs) proposed by Behler and Parrinello[68-70] (Figure 3(a)). They began by training a NN model on lower-accuracy DFT data sets (named ANI-1x). ANI-1x was constructed by using active learning and contained DFT data for 5M conformations of molecules with an average size of 15 atoms. Then they retrained a TL model (ANI-1ccx for short) on a much smaller data set (about 500k intelligently selected conformations from ANI-1x) at CCSD(T)/CBS level of accuracy (the coupled cluster considering single, double, and perturbative triple excitations calculations are combined with an extrapolation to the complete basis set limit (CBS)). TL was implemented by copying and fine-tuning the parameters of the first model. The obtained general-purpose potential, ANI-1ccx, and data set exceed the accuracy of DFT in benchmarks for isomerization energies, reaction energies, molecular torsion profiles, and energies and forces at non-equilibrium geometries, while being roughly nine orders of magnitude faster than DFT. To generalize the deep learning network to the new single-molecule fluorescence microscopy (SMFM) system, Li et al. developed an SMFM trajectory selector to improve the sensitivity and specificity of DNA point mutation detection based on single-molecule recognition by applying transfer learning[71], the biggest advantage of introducing transfer learning is that it does not require large training datasets. Similarly, for small datasets, Li proposed an approach named "Molecular Prediction Model Fine-Tuning (MolPMoFiT)" based on transfer learning, including self-supervised pre-training and task specific fine-tuning[72]. They pre-trained on one million bioactive molecules from ChEMBL and fine-tuned for some Quantitative structure property/activity relationship (QSPR/QSAR) tasks. Although the amount of data in the molecular field has grown dramatically over time, modeling and prediction of smaller datasets is still one of the common problems. Applying transfer learning is therefore an important approach not only to overcome the challenges of small datasets but also to benefit from the rich data contained in the publicly available datasets.

Organic synthesis is usually a complex process. The information about the organic molecules and functional groups involved in most organic reactions can be applied to construct deep learning models, but the prediction of organic reactions is still a great challenge when considering the complexity of the functional group environment. As shown in Figure 3(b), Pesciullesi et al. challenged the Molecular Transformer model to predict reactions on carbohydrates, where region and stereoselectivity are notoriously difficult to predict[73]. They used a dataset containing twenty thousand carbohydrate reactions from the literature, including protection/deprotection and glycosylation sequences. They explored multi-task learning and transfer learning and demonstrated that adapting the Molecular Transformer, called Carbohydrate Transformer, achieved significantly better performance than general models for carbohydrate transformation and models trained specifically on carbohydrate reactions. Their work suggests that transfer learning can generate a specialized model with high accuracy for predicting carbohydrate reactions. They validated these predictions through synthetic experiments involving region-selective protection and stereo-selective glycosylation of lipid-linked oligosaccharides. The transfer learning approach should be applicable to any desired reaction class.

*Polymer*

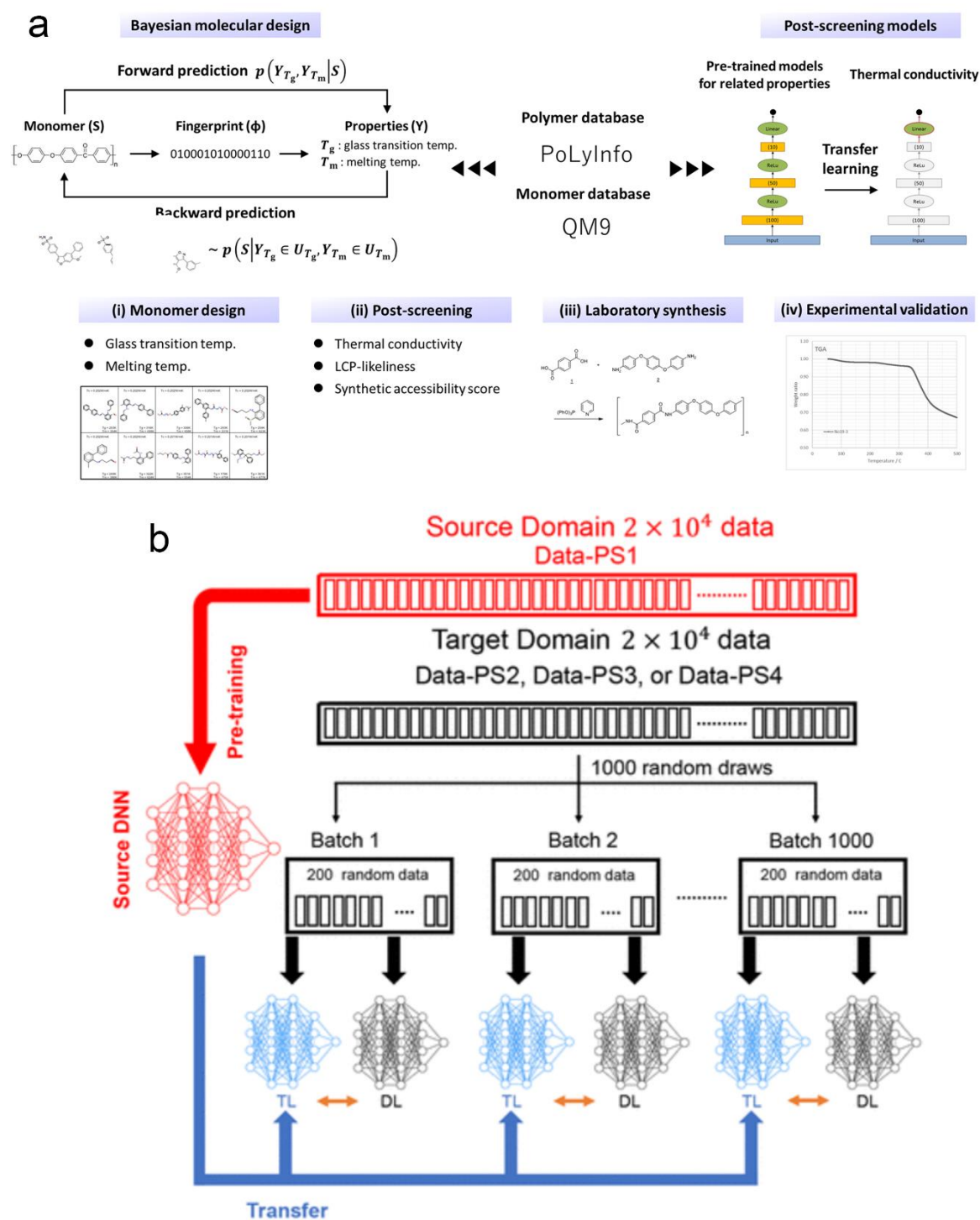

**Figure 4** (a) ML&TL -assisted de novo design and experimental validation of new polymers. Reprint from Ref [76] (b) A schematic of the procedure for testing the performance of transfer learning from source domain (Data-PS1) to target domain (Data-PS2, Data-PS3, or Data-PS4). Reprint from Ref [77]

Small organic molecules are varied, with a potential chemical space of $10^{60}$ structures, yet only a maximum of $10^8$ compounds have been reported and the development is still in the exploratory stage[74, 75]. Exploring such a vast structural space through traditional trial-and-error experimental methods is still challenging. The emergence of machine learning methods has provided efficient ways to expand the search space of organic compounds. However, there are still obstacles to the application of machine learning in the field of polymer design, mainly due to the limited property data of polymers, the gap between theoretical design and experimental synthesis, and the divergence between expert knowledge and the information learned by machine learning models. Wu *et al.* addressed the problem of small dataset on polymer thermal conductivity (only 28 training instances) by introducing a transfer learning framework and adopting proxy properties (glass transition temperature ($T_g$) and melting temperature ($T_m$)) similar to the target property as the design goal (Figure 4(a)).[76] They pre-trained the model on the alternative properties in a large dataset gained from PoLyInfo and QM9 to learn features related to the thermal conductivity of the polymer, which were then used on the small dataset to accurately predict the thermal conductivity. The best transferred models performed a mean absolute error (MAE) of 0.0204 W/mK, a 37% improvement in prediction accuracy compared to direct prediction model. Moreover, the negative coefficient of determination of direct prediction indicates the model's unreliability, which is due to the limited dataset size. Three polymers named polyamides 4, 13 and 19, were selected from 1000 candidate structures with reference to the ease of synthesis of the polymers and then experimentally prepared, and the three polymers obtained had thermal conductivities in the range of 0.18~0.41 W/mK. This study illustrates the common application way of transfer learning in the field of polymers, specifically, pre-training and fine-tuning based on larger datasets to enhance the performance of ML models, which reflects transfer learning's core of "knowledge sharing". Data on the properties of polymers for various applications are also extremely scarce. Shi et al.[77] use a similar framework combining deep neural network (DNN) and transfer learning to construct predictions of adhesive free energies between polymer chains with a defined sequence and patterned surfaces, while taking into account the impact of database size on the final performance of the models (Figure 4(b)). The source domain dataset contains $2 \times 10^4$ polymer sequence and adhesive free energies, and a full DNN is trained on this large dataset and its weights are retained for subsequent fine-tuning of the target domain models. This transfer learning strategy is applied to three different target domains, corresponding to three different patterned surfaces. The results showed that the accuracy of the direct prediction model is generally lower than that of the transfer learning model, regardless of the size and type of the target domains, e.g., for the Data-PS2 dataset containing 200 data, the coefficient of determination ($R^2$) improved from –0.0089 ± 0.1956, which is almost a random-guessing accuracy, to 0.8303 ± 0.0747. The performance of the model is dramatically improved, and the transfer learning model showed greater stability than the direct learning model due to its reduced susceptibility to random selection and other disruptive factors.

*Biomacromolecules*

Biological macromolecules, such as proteins, enzymes, DNA and RNA, are important building blocks of life and play an important role in various biological activities and metabolism. In recent years, AI models have made important breakthroughs in the field of proteins[78]. For example, the introduction of AlphaFold[21] and RoseTTAFold[79-81] opens up a new way to solve the centuries-old problem of protein folding by predicting three-dimensional protein conformations from amino acid sequence information. However, the success of these predictions has been achieved through an end-to-end model, in which the spatial structure of macromolecules is predicted from sequences in a "black box", and the lack of dynamic folding processes has greatly limited the understanding of the physical processes of protein folding.

Molecular dynamics (MD) simulation is the main method for protein folding, stabilization, and interaction. Its accuracy and applicability mainly depend on an advanced force field and efficient global energy minimum search engine. However, the high computational cost of high-level quantum mechanical (QM) method and the complexity of large proteins bring great challenges to the establishment of ML force fields (MLFFs) of large proteins. As shown in Figure 5(a), Han *et al.* designed an inductive TL force field (ITLFF) protocol that can build protein force fields in seconds[33]. The ITLFF constructs a force field (energy and atomic forces) by dividing proteins into 21 residue-based fragments (20 capped residues and one cap, fragment-based QM method). For each fragment, they first trained a source NN model based on 20,000 data calculated by low-level QM methods (ωB97XD/6-31G*), resulting in a holistic root-mean-squared error (RMSE) of loss function of 0.090. Then, they calculated 1,000 data by using double-hybrid functional (DSD-BLYP/def2-TZVPP with D3BJ dispersion correction) and trained a target NN model with a holistic RMSE of loss function of 0.051. Validated by 18 proteins, the force fields built by ITLFF achieve considerable accuracy with a mean absolute error (MAE) of 0.0039 kcal/mol/atom for energy and an average RMSE of 2.57 kcal/mol/Å for atomic forces. High-performance ITLFF provides a broad prospect for accurate and efficient protein dynamic simulation. Theoretically, ITLFF can also be applied to a variety of problems in biology, chemistry, and material science. Further, a TL-based deep learning (TDL) model for effective full quantum mechanics calculations (TDL-FQM) was developed by using the different fragment-based QM methods (two- or one-residue fragments). Validated by 15 proteins, the force fields built by TDL-FQM achieve considerable accuracy with a MAE of 0.01 kcal/mol/atom for energy and an average RMSE of 1.47 kcal/mol/Å for atomic forces. This strategy also demonstrates the scalability of TL methods in fragment-based QM approaches.

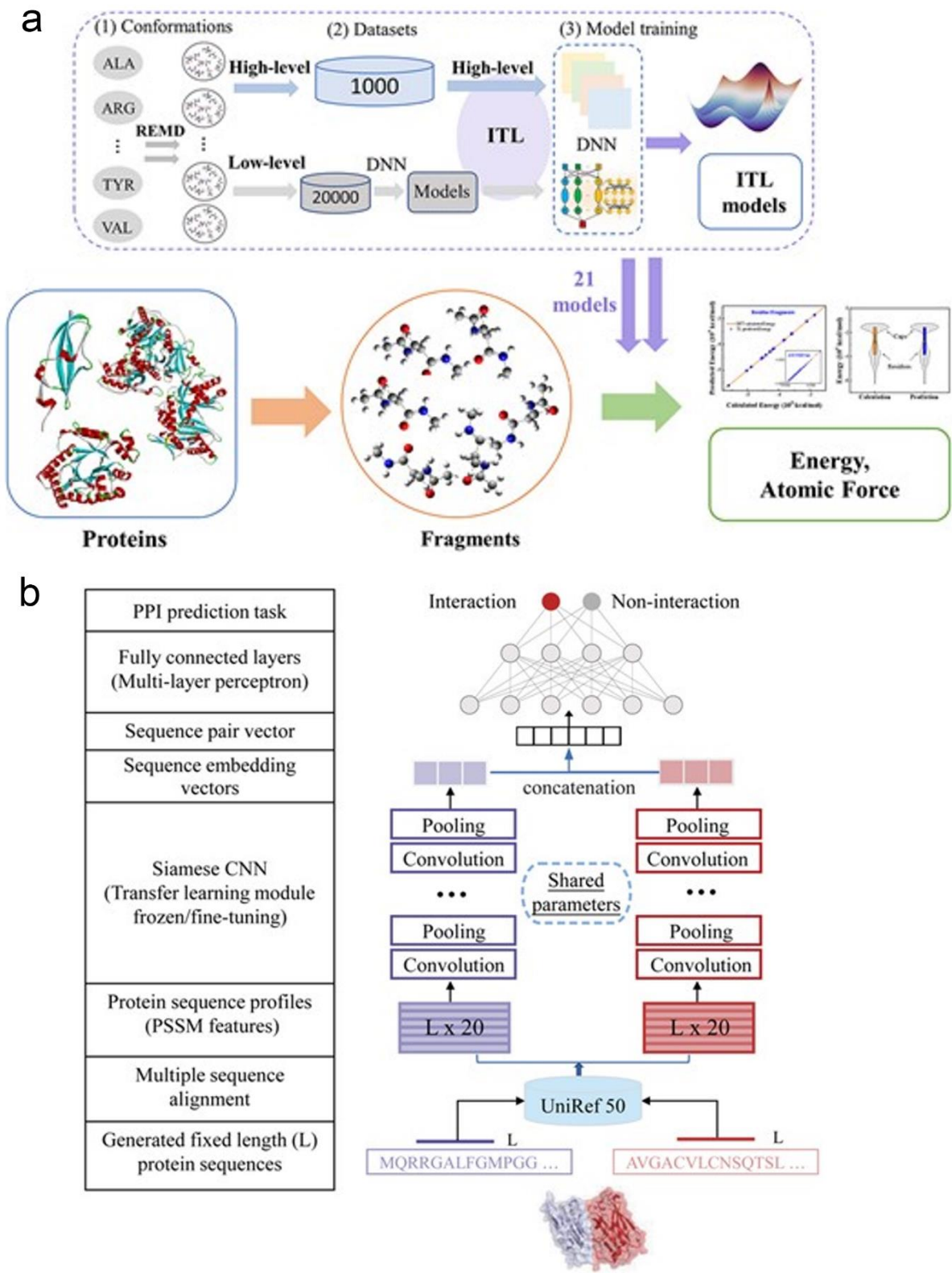

**Figure 5** (a) ITLFF architecture, where proteins are divided into a series of fragments, which are used for constructing the universal protein force field and calculating the energy and atomic forces. The energy and atomic forces of fragments are calculated using the 21 fragmental models trained with inductive transfer learning. Reprint from Ref [33]. (b) Siamese CNN module and a prediction module. Reprint from Ref [82].

Apparently, fragment-based QM approaches, NN models and TL methods can be effectively combined. It is worthwhile to be used in the force field construction of biological macromolecules such as DNA/RNA. In addition, the level of precision of QM approaches can also be obtained on demand by modifying the computational precision of the data in the target model.

Improving the accuracy of theoretical simulations of biomacromolecules through transfer learning can facilitate the understanding of the physical processes of protein folding. At the same time, transfer learning can also be applied to the prediction of protein interactions. Yang *et al.* used evolutionary sequence profile features and a Siamese convolutional neural network (Siamese CNN) framework for predicting human-virus protein-protein interactions (PPIs), mainly consisting of a module for pre-acquiring protein sequences and a module containing CNN and MLPs[82]. The main modules are shown in Figure 5(b). They assessed the accuracy of a series of PPIs predictions through 5-fold cross-validation. The accuracy of the deep learning model is low for training sets with small data sizes. Notably, they introduced two transfer learning methods: frozen and fine-tuning type, respectively. After the application of the two transfer learning methods, the prediction performance is improved by 50% on average, and the AUPRC of the frozen type transfer learning is better than that of the fine-tuning type transfer learning method. Finally, they employed the frozen type of transfer learning to predict human SARS-CoV-2 PPIs, demonstrating the effective application of prior knowledge obtained from a large source dataset/task to a smaller target dataset/task, thereby enhancing prediction performance.

*Inorganic compounds*

Inorganic compounds are one of the core and foundations of modern information society, and play an important role in photovoltaic systems, integrated circuits, spacecraft technology, lighting applications, catalysts, solar materials, superconductors, inorganic thin films, and other fields. ML methods have been utilized to develop and predict the properties of advanced inorganic materials effectively, surpassing the limitations of trial-and-error experiments and traditional DFT calculations. This stimulates new ways and advancements in material informatics of inorganic compounds.

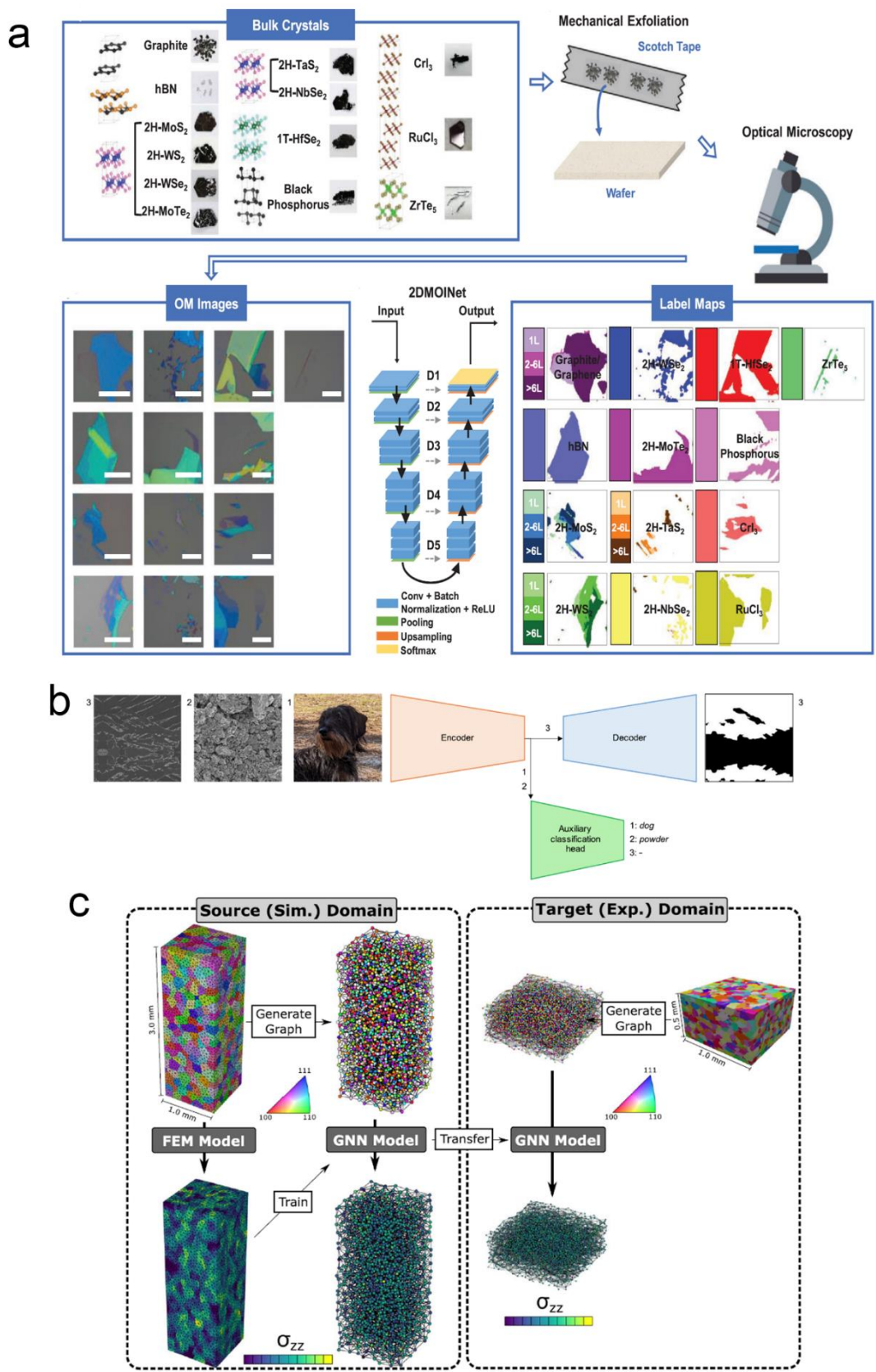

**Figure 6** (a) The flow chart of the proposed deep learning based optical identification method. Reprinted from Ref. [83]. (b) Summary of the pre-training and fine-tuning procedure. Reprinted from Ref. [84].

(c) Overview of transfer learning approach for GNN surrogate model evaluation. Connectivity of the various efforts used to evaluate the accuracy of GNN surrogate models for predicting the stress response in individual grains during elastic loading. Reprint from Ref. [86].

Advanced microscopes and/or spectroscopic tools play an indispensable role in experimental materials research, as they provide rich information about material processes and properties. However, the interpretation of imaging data often heavily relies on the "intuition" of experienced researchers. As a result, many deep graphical features obtained through these tools remain unused due to the challenges in data processing and finding correlations. As shown in Figure 6 (a), Han et al.[83] demonstrated a neural network-based algorithm called 2D Material Optical Identification Network (2DMOINet), using the optical characterization of two-dimensional materials as an example. 2DMOINet exhibits high prediction accuracy and real-time processing capability. Further analysis showed that 2DMOINet can extract deep graphical features such as contrast, color, edges, shapes, sheet sizes, and distributions. Based on these features, an integrated approach was developed to predict the most relevant physical properties of 2D materials. Furthermore, they demonstrated that the trained 2DMOINet can adapt to different applications through transfer learning. The basic idea is to use the pre-trained 2DMOINet as the initialization for the new training problem, rather than random initialization. Through this approach, they can train 2DMOINet to address new optical identification/characterization problems while achieving a good balance between prediction accuracy and computational/data costs. Hundi et al.[84] simulated the microstructural damage of hexagonal boron nitride (h-BN) at different radiation levels and temperatures to predict its residual strength from the final atomic positions (Figure 6(b)). They employed models such as convolutional neural networks and multi-layer perceptrons to predict structure-property mappings. By developing low-dimensional physical descriptors to statistically describe defects, they showed that a microstructural representation tailored for specific purposes can achieve good prediction accuracy at lower computational costs. Furthermore, using transfer learning, they also explored the adaptability of trained deep learning agents in predicting the structure-property mappings of other 2D materials. The results indicated that to achieve good prediction accuracy (≈95% $R^2$), the initially trained agent ("learning from scratch") required 23-45% of the simulation data, while the agent adapted to different materials ("transfer learning") only needed approximately 10% or less. This suggests that transfer learning could be a potential game-changer in materials discovery and characterization methods. Recently, automated quantitative analysis of microstructural constituents has been achieved through deep learning methods. However, their drawbacks include data efficiency, poor domain generality across datasets, inherent trade-offs between the cost associated with expert-annotated data and the wide diversity of materials. To address these two challenges, Goetz et al.[85] propose the use of a subclass of transfer learning methods called "Unsupervised Domain Adaptation" (UDA). UDA tackles the task of finding domain-invariant features when provided with annotated source data and unannotated target data, optimizing the performance of the latter.

In predicting macroscopic mechanical properties, the predictive accuracy of ML models has already surpassed that of traditional mean-field theories (models), with the introduction of TL methods, accurate models can be obtained even under scarce datasets. As shown in Figure 6(c), Pagan *et al.*[86] applied Graph Neural Networks (GNN) to predict the elastic response of two alloy systems, Low solvus high-refractory (LSHR) Ni superalloy and Ti-7 wt% Al (Ti-7Al), representing cubic and hexagonal elastic responses, respectively. They developed and trained two GNN proxy models (Gaussian Mixture Convolution) to predict the average elastic response of LSHR and Ti-7Al grains (components of grain stress tensor along the loading direction) and tested them using crystal elasticity finite element method (CEFEM) deformation simulations that explicitly considered grain microstructures. During training, the predicted accuracy was compared with predictions from traditional mean-field theories and retained CEFEM simulations. Then, a transfer learning strategy was applied, and the GNN models trained with CEFEM simulations were transferred to a separate data domain to predict the average elastic response of polycrystalline microstructures measured using near-field high-energy X-ray diffraction microscopy (HEDM).

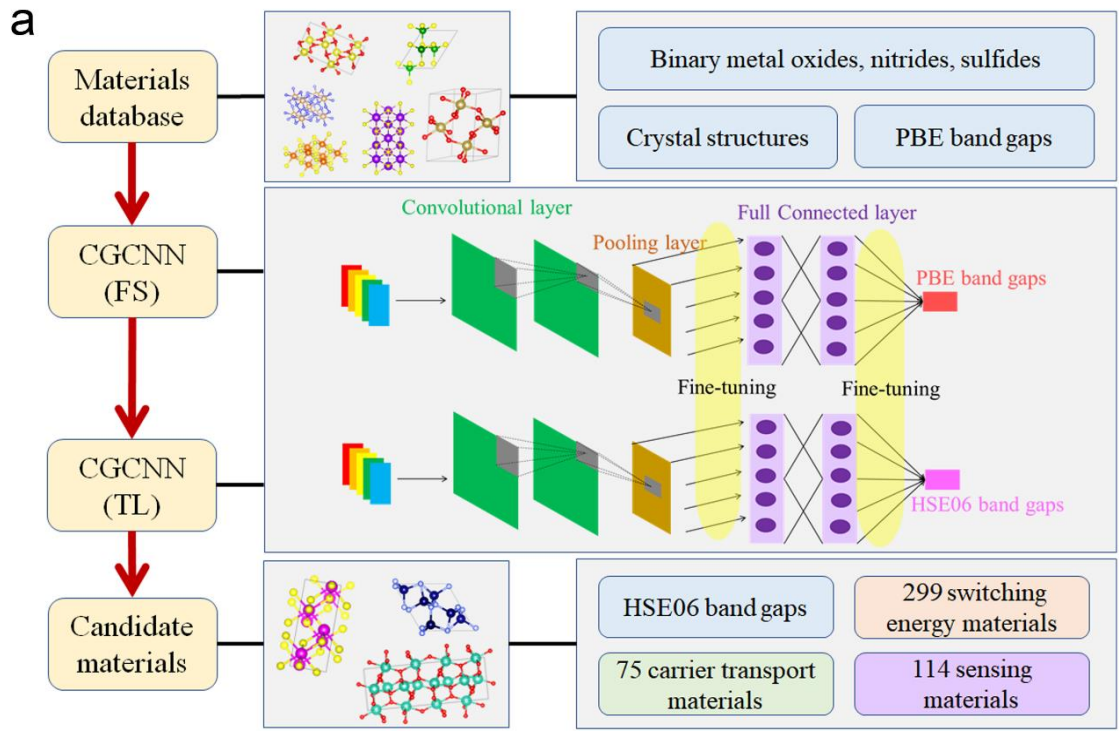

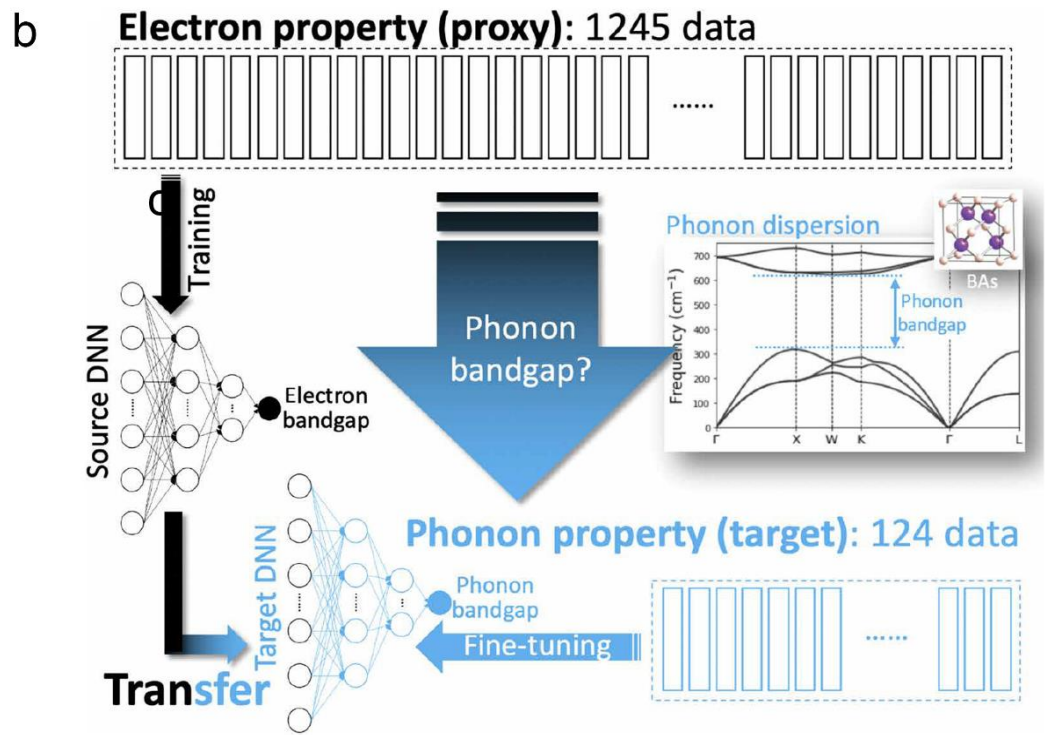

**Figure 7** (a) Framework of CGCNN-TL. Reprinted from Ref. [32]. (b) Schematic of TL from electron property to phonon properties. Reprinted from Ref. [87].

Band gap is widely recognized as one of the fundamental electronic properties of materials. Unfortunately, the long period of band gap prediction methods based on experimental and high-throughput calculations has hindered the development of new materials. Using high-precision functional (*e.g.*, hybrid functional, heyd-scusera-ernzerhof (HSE06)) to calculate the band gap can get similar results to the experimental measurement, but the calculation cost is often very high. Although the calculation using Perdew-Burke-Ernzerhof (PBE) functional reduces the cost, it greatly underestimates the actual band gap value. Wang *et al.* developed a TL model (named CGCNN-TL) based on the crystal graph convolutional neural network (CGCNN) model[32] (Figure 7(a)). They first trained a source model (called CGCNN-FS) that can predict the PBE band gap based on 1,503 data taken from the Materials Project database, resulting in an MSE of 0.35 eV and an $R^2$ of 0.89. Then, they compiled a dataset that contains 64 HSE06 band gaps by high-throughput calculations (less than 5% of the 1503 data) and established a TL model for rapidly predicting the HSE06 band gap, with an MSE of 0.21 eV and an $R^2$ of 0.98. The well-trained hyper-parameters of CGCNN-FS model are the initialization of the CGCNN-TL model, and the hyper-parameters inside the convolutional layer, pooling layer and fully-connected layer were fixed (that is, do not change with iteration). In the training process, only the hyper-parameters between the pooling layer and the first fully-connected layer, the second fully-connected layer and the output layer were fine-tuned. This work demonstrates the feasibility of TL in band gap prediction, from the low-level PBE to high-level HSE06.

Electronic properties are usually more readily available than phonon properties. The ability to use electronic properties to help predict phonon properties can greatly benefit material design applications such as those in thermoelectrics and electronics. Liu et al.[87] demonstrated the capability of transfer learning (TL) using a multilayer perceptron-based model (Figure 7(b)), where knowledge learned from training a machine learning model of the electronic bandgap of 1245 semiconductors is transferred to an improved model trained using only 124-data, and the resulting model can be used to predict a variety of phonon properties (phonon bandgap, group velocity, and heat capacity). The average absolute error of TL's predictions for the three phonon properties was reduced by 65%, 14% and 54%, respectively, compared to the directly trained model. The results also show that TL can take advantage of less accurate proxy attributes to improve the target attribute model as long as they encode the composition-attribute relationship, which is a distinctive feature of materials informatics. For example, in order to develop materials for new applications where experimental data are sparse and difficult to collect, one can apparently use calculations (e.g., coarse-grained molecular simulations) to quickly generate large amounts of data as proxy labels, which can then be used to construct predictive models using TL approach.

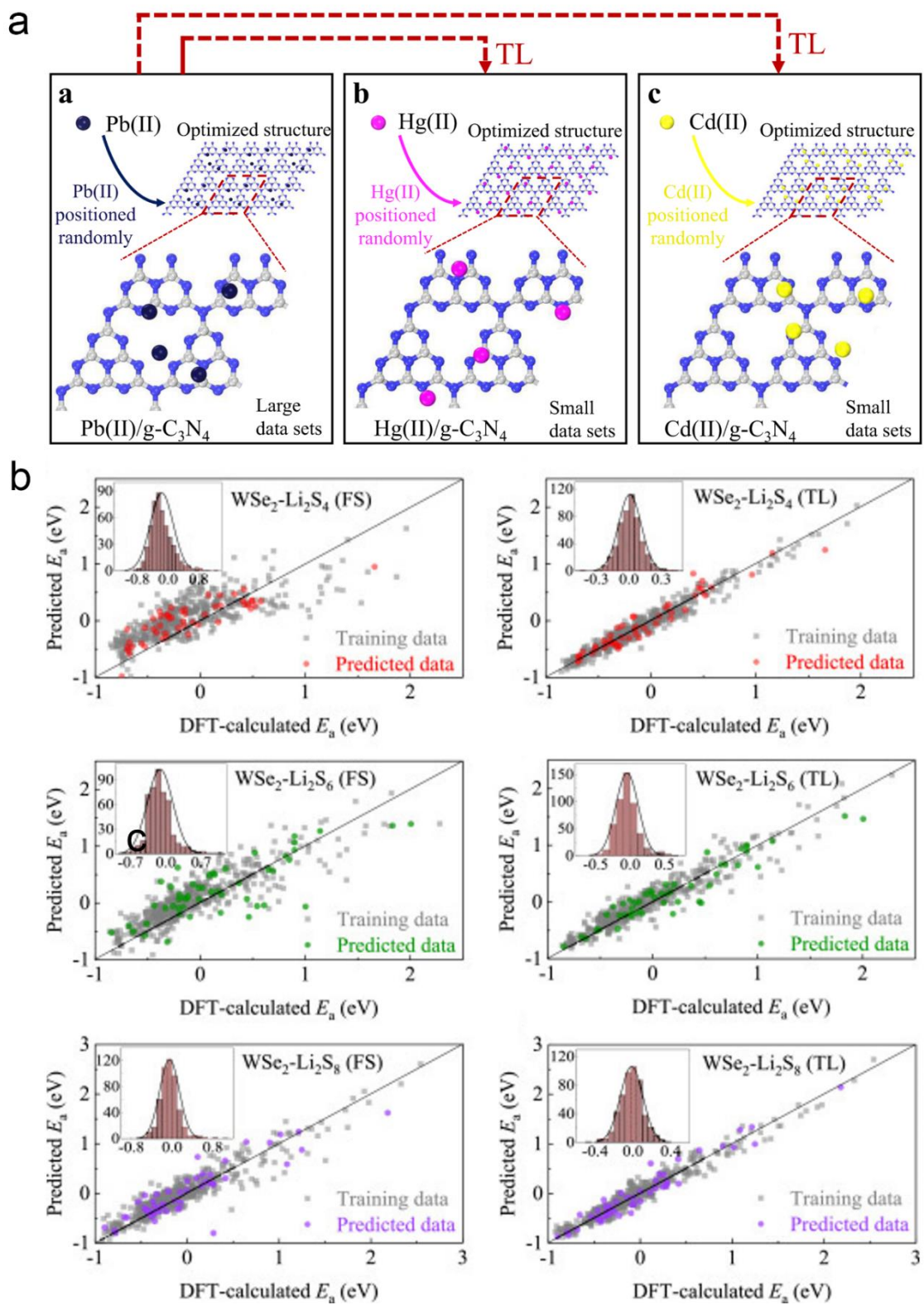

**Figure 8** (a) Schematic model of adsorption process towards HMIs on the surface of g-$C_3N_4$. Reprint from Ref. [31]. (b) Correlation plots of binding energy against DFT and ML, along with histograms of error distributions between DFT and ML. Results on the left were obtained by FS training, whereas results on the right were obtained by TL training. Reprint from Ref. [96].

Besides the electronic structural property prediction, the prediction of adsorption energy is also a challenge for materials. The calculations of adsorption energy are needed in many fields of research. For example, the decontamination ability of pollutant adsorbent[31, 88], the multi-step transformation in surface catalytic reaction[89-92], the diffusion of lithium ions on the electrode in lithium-ion batteries (LIBs)[93], and the adsorption and transformation of polysulfide (LiPS, $Li_2S_x$, $x$ = 2, 4, 6, 8) by cathode materials in lithium-sulfur (Li-S) batteries[94, 95]. Adsorption ability of an adsorbent relies on the active sites and the corresponding activity intensities, which is currently hardly detectable. Theoretical prediction provides another way to understand the mechanism of adsorption process and explore the efficient adsorbent. However, the configuration space provided by a wide variety of materials and the complex relationship between adsorbent active sites and activity strength suggest that traditional structural optimization methods based on inherently time-consuming ab initio calculations are particularly challenging. As shown in Figure 8(a), Wang *et al.* developed a TL model for predicting the adsorption energies at arbitrary sites accurately and quickly, based on deep NN. They chose 2D g-$C_3N_4$ material and heavy metal ions ($Pb^{2+}$, $Hg^{2+}$, $Cd^{2+}$) as a case study[31]. The adsorption energies of $Pb^{2+}$ on the g-$C_3N_4$ were evaluated by 7,000 single-point DFT calculations, and the predictive model was established by using from scratch training, with a RMSE of 0.043 eV. The adsorption energies of $Hg^{2+}$ and $Cd^{2+}$ were evaluated by only 700 DFT calculations, and the predictive model was established by using TL training, with RMSEs of 0.012 eV and 0.051 eV, respectively. Combined with TL, the overall adsorption strength predicted was $Cd^{2+}$ > $Hg^{2+}$ > $Pb^{2+}$, which was consistent with the experimental results.

Zhang *et al.* applied this TL-based adsorption energy prediction method to the adsorption energy of polysulfide in the cathode material of lithium sulfur battery[96]. As a case study, layered compound $MoSe_2$ was selected as a cathode material to adsorb the LiPS. The predictive NN model was trained on a total of 9,800 data, showing six orders of magnitude faster than the conventional DFT calculations, with a low MAE of 0.1 eV. Based on the TL strategy, they demonstrated that the NN model can be transferred to other layered compounds with a similar $AB_2$ structure to $MoSe_2$, and can efficiently predict their adsorption strengths with hosts. $WSe_2$ was employed as a case to validate the TL method based on a total of 1,500 data, with a low MAE of ~0.12 eV. Then, they concluded that $MoSe_2$ had a stronger adsorption strength than $WSe_2$ when adsorbing the LiPS, and only one-seventh of the NN training data was required. The compared models of from scratch (FS) training and TL are shown in Figure 8(b). This work provides important technical support for screening the cathode materials of Li-S batteries to better inhibit the shuttle-effect. Further, the authors then applied this method to the systems of $VTe_2$-$Li_2S_6$ and $FeI_2$-$Li_2S_6$, to predict the adsorption energies at the arbitrary sites. They found that $Li_2S_6$ can be strongly adsorbed by the $VTe_2$ against solvent adsorption, demonstrating that $VTe_2$ is a promising cathode material to suppress the shuttle-effect and improve the cyclic stability of Li-S batteries.

A machine learning model was developed to predict the energy of materials using their compositions and structures.[97] The authors first compiled a large database containing 2 million DFT calculation results, including data points from online databases and their own calculations obtained through compatible parameters. Despite the large scale of this dataset, it was somewhat biased due to many calculations being focused on relatively few different crystal prototypes. To mitigate this issue, the authors proposed a TL approach where a general-purpose model is retrained for specific crystal structures. They demonstrated this idea with an experimental case of a quaternary perovskite, showing that transfer learning can speed up the model's training process two-fold. By excluding V-containing compounds from training, they also demonstrated that the network can reliably extrapolate to the unknown regions of the periodic table.

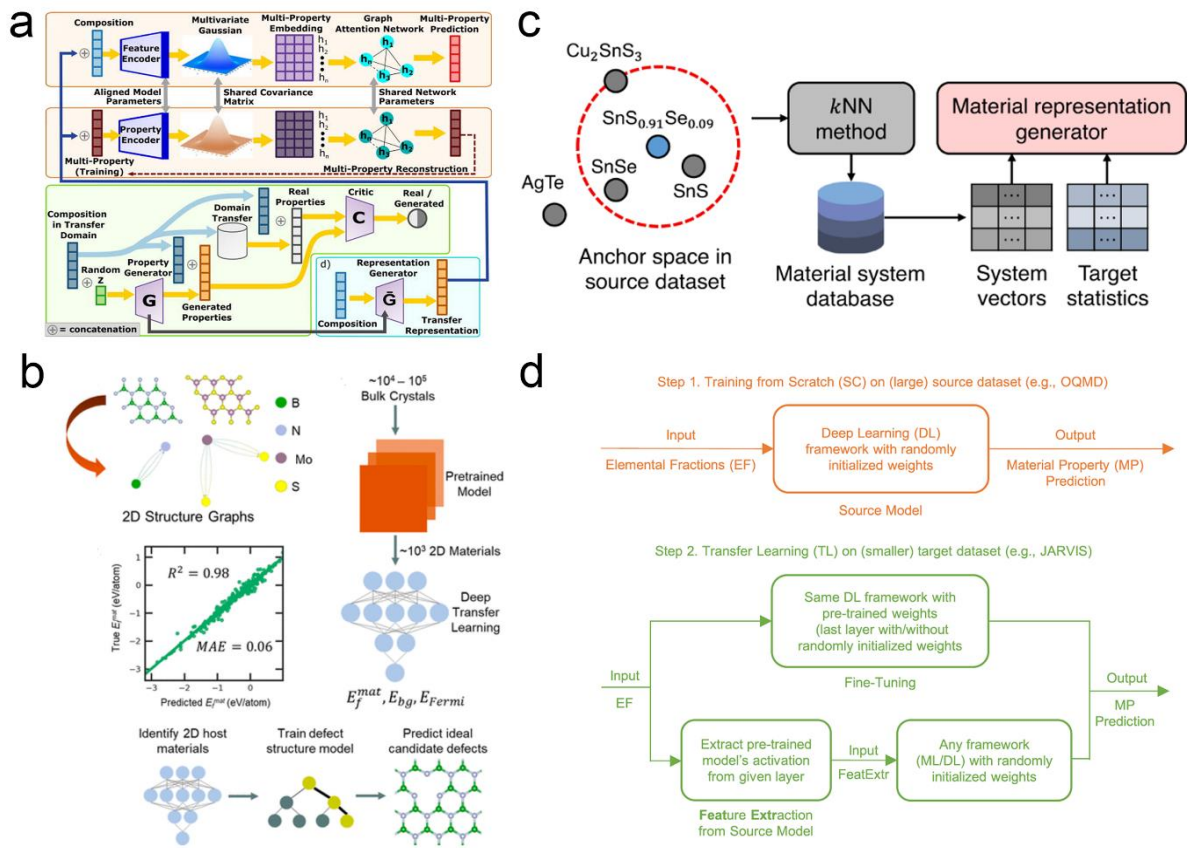

**Figure 9** (a) The H-CLMP(T) framework. Reprint from Ref [98]. (b) Deep transfer learning property prediction for 2D materials. Reprint from Ref [105]. (c) The overall process of system-identified material descriptor (SIMD) to generate the system-identified features of the input chemical composition in the transfer learning environments. Reprint from Ref [106]. (d) The proposed cross-property deep-transfer-learning approach. Reprint from Ref [107].

The rational design of advanced functional materials often also faces the difficulty of lacking available training data. Kong et al.[98] proposed a model called "Hierarchical Correlation Learning" for multi-property prediction (H-CLMP), which is based on hierarchical correlation learning. They used transfer learning (H-CLMP(T)) to learn chemical interactions from other domains to discover materials with unique optical properties (shown in Figure 9(a)). The prediction task involved measuring the optical absorption coefficients of ten photon energies, spanning the visible spectrum, and extending into the ultraviolet, for a complex metal oxide material containing three cation elements (excluding oxygen). The model consisted of four highly interdependent components: (a) a multi-property prediction model that takes combined and transformed data as input, (b) a target property autoencoder trained jointly with (a), (c) a separately trained generator for transfer learning, and (d) the deployment of the generator for transfer learning. Components (a) and (b) formed the core H-CLMP model, while components (c) and (d) constituted its transfer learning extension, H-CLMP(T). H-CLMP also integrated prior scientific knowledge through transfer learning, using physics-based computational data from materials projects to encode material chemistry in the generative model and enhance multi-property prediction. In a new compositional space where no training data was available, H-CLMP outperformed state-of-the-art methods in predicting solar material optical absorption characteristics.

Two-dimensional (2D) materials are one of the most important types of functional materials today, and the wealth of modifications associated with 2D materials is continuing to expand the compound space[99-103]. Defects in materials are critical to the performance impact, and thus defect engineering is a commonly used method for performance modulation.[104] Machine learning methods are able to efficiently mine and understand atomic-scale defect conformational relationships from the vast array of defect types and influencing factors. Frey et al.[105] combined deep learning, transfer learning and ab initio materials design to systematically investigate nearly a hundred 2D materials (both van der Waals and non-van der Waals) and rapidly identify the most promising defect structures for quantum emission and neuromorphic computing. The transfer learning procedure is shown in Figure 9(b). Starting from 4000 2D materials, they used transfer learning to leverage models trained on tens of thousands of bulk crystal structures for deep learning-driven prediction of 2D material properties, which is crucial for identifying promising defect host systems. A graph network model was pre-trained on a large dataset of bulk crystals, and transfer learning was used to fine-tune the model weights for predicting the formation energy, bandgap, and Fermi energy of 2D materials. The formation energy of 381 materials predicted by deep transfer learning was compared to DFT calculations, demonstrating the $R^2$ value and MAE.

Thermoelectric materials as energy harvesting devices and generators have attracted widespread attention. However, discovering new high-performance thermoelectric materials with alloys and dopants is challenging due to the structural diversity and complexity. To facilitate efficient data-driven discovery of novel thermoelectric materials, Na[106] constructed a public dataset that includes experimentally synthesized thermoelectric materials and their measured thermoelectric properties. For the collected dataset, they were able to build predictive models with $R^2$ regression scores greater than 0.9 to predict experimentally measured thermoelectric properties from the material's chemical composition. Additionally, to make the models transferable, they employed a transfer learning approach. The source model consisted of 215,683 thermoelectric materials, and through transfer learning and proposed material descriptors (system-identified material descriptor, SIMD), they significantly improved the $R^2$ score in predicting the experimental figure of merit (ZTs) of materials from completely unexplored material groups, increasing it from 0.13 to 0.71. Figure 9(c) shows the process of SIMD that they devised for the generation of system-identified feature in the transfer learning environments.

As shown in Figure 9(d), a deep transfer learning framework for cross-material property prediction, which enables transferring knowledge learned by predictive models trained on large datasets to construct predictive models on smaller datasets of other target properties[107]. The main advantage of the proposed cross-property deep transfer learning framework is that it allows the development of powerful and accurate models on smaller property datasets, which may not have sufficient data available for training from scratch. This transfer learning framework consists of two steps: first, training a deep learning model on a large dataset, and second, using the source model to build a target model for the target property on a smaller target dataset. This can be achieved by either fine-tuning the source model on the target dataset or using the source model as a feature extractor. In addition to prediction errors, this predictive model inherits discrepancies between DFT calculations and experimental measurements. Jha et al. addressed this challenge by demonstrating that using deep transfer learning, existing large-scale DFT calculation datasets, such as the Open Quantum Materials Database (OQMD), can be combined with other smaller DFT calculation datasets and available experimental observations to build robust prediction models. They developed a highly accurate model for predicting the formation energy of materials from their compositions. Using a dataset of 1643 observed data points, the method achieved an average absolute error (MAE) of 0.07 eV/atom, significantly outperforming existing DFT-based machine learning (ML) prediction models and comparable to the MAE of DFT calculations themselves.

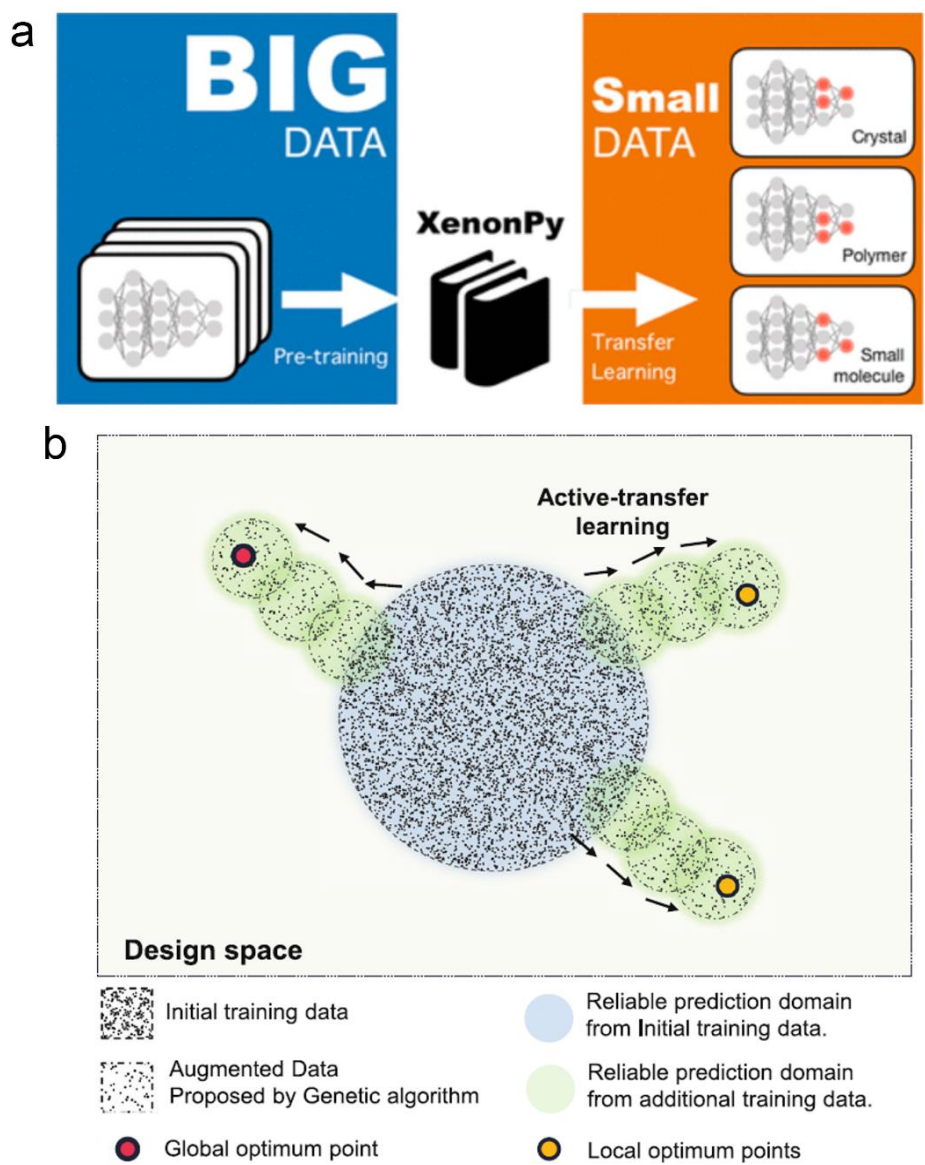

**Figure 10** (a) XenonPy.MDL model library, has the great potential of transfer learning to break the barrier of limited amounts of data in materials property prediction using machine learning. Reprinted from Ref [26] (b) Schematic of deep learning framework for material design space exploration. Reprinted from Ref [108]

As shown in Figure 10(a), to facilitate the widespread use of TL methods, Yamada et al. developed a comprehensive library of pre-trained models, named XenonPy.MDL, which was fed diverse sets of material property data into NNs and other ML models (such as RF model)[26]. XenonPy.MDL contains many molecule property prediction models trained on QM9 data (12 common properties of 133805 small organic molecules), such as density, viscosity, melting temperature, heat capacity, thermal conductivity, etc.

Neural network-based generative modeling has been actively investigated as an inverse design method for finding new materials in a wide design space. However, the applicability of traditional generative models is limited because they cannot access data outside the scope of the training set. Advanced generative models designed to overcome this limitation also suffer from the weak predictive power of the unseen domain. Kim et al.[108] proposed a deep neural network-based forward design method that effectively searches for high-quality material far beyond the domain of the initial training set (Figure 10(b)). This approach compensates for the weak predictive ability of neural networks over unseen domains by incrementally updating the neural network using active transfer learning and data augmentation methods. The authors use a genetic algorithm to generate candidate material structures that are predicted by the DNN until the DNN converges. After calculating the real properties of the candidate materials using finite element simulations, data augmentation techniques (replicated 50 times and divided 9:1 into training and validation sets) were used to add to the training data to increase the density of the newly added data. Whereas DNN training is achieved by reducing the learning rate and re-training using transfer learning techniques, this way the required training time is significantly reduced.

## 4. Frontiers in transfer learning

ML in the material field is still facing various challenges of training on small datasets, and the advantage of knowledge reuse of TL, despite solving a large number of small dataset problems, will be in this predicament for a long time in the future, due to the uneven level of data management and the different speeds of development of various materials.

Lack of sufficient data is still the main factor for the machine learning dilemma in the materials field, and the framework of transfer learning strategy to solve the problem is to pre-train on existing data-rich datasets, fine-tune on small datasets, and do meta-learning on the rich data at the same time to overcome the data scarcity problem in the target domain. In the process of applying a TL strategy, the amount of data and feature distributions of the source domain and the target domain often have large differences, or even the target domain has no (or only a very small number of) labeled samples, in which case the models pre-trained based on the source domain are not able to be successfully applied to the target domain and enhance the performance of TL model. Domain adaptation will be used to address such challenges. Moreover, when the data types of the pre-training dataset and the target dataset are quite different (e.g., the pre-training data is images and the data in the target dataset is text), multi-step domain adaptation is required. This is not negligible for the study of material or molecule design, which translates into the specific problem of the difference in the form of experimental and computational data. Although ML methods are much more efficient compared to experimental methods or high-throughput calculations, it is not cost-free. The collection of data and multiple trainings of the model have time and money costs. Thus, TL models with high generalization capability should be constructed to reduce costs and extend to a wide range of material or molecule types.

Transfer learning, as a potent machine learning strategy, has fostered the advancement of various inter, and multidisciplinary fields, while also benefiting from their progress in a synergistic manner. In the future, TL-based ML will continue to revolutionize materials science.

## Conflict of Interest

The authors declare no financial or non-financial competing interest.

## Acknowledgements

The authors are grateful for the financial support provided by the National Key R&D Program of China (No. 2021YFC2100100), and the Shanghai Science and Technology Project (No. 21JC1403400).## Reference

1. Fischer, C. C.; Tibbetts, K. J.; Morgan, D.; Ceder, G. Predicting crystal structure by merging data mining with quantum mechanics, *Nat. Mater.* **2006,** *5* (8), 641-646.
2. Byrd, E. F.; Rice, B. M. Improved prediction of heats of formation of energetic materials using quantum mechanical calculations, *J. Phys. Chem. A* **2006,** *110* (3), 1005-1013.
3. Rice, B. M.; Hare, J. J.; Byrd, E. F. Accurate predictions of crystal densities using quantum mechanical molecular volumes, *J. Phys. Chem. A* **2007,** *111* (42), 10874-10879.
4. Li, J.; Sode, O.; Voth, G. A.; Hirata, S. A solid–solid phase transition in carbon dioxide at high pressures and36